\documentclass[proof]{WileyASNA-v1}
\usepackage{graphicx}

\articletype{Article Type}%

\received{}
\revised{}
\accepted{}

\def\psp{$P_{\rm spin}$}
\def\ltsima{$\; \buildrel < \over \sim \;$}
\def\lsim{\lower.5ex\hbox{\ltsima}}
\def\gtsima{$\; \buildrel > \over \sim \;$}
\def\gsim{\lower.5ex\hbox{\gtsima}}
\def\XMM{\textit{XMM--Newton}}
\def\lx{$L_{\rm X}$}
\def\lum{\mbox{erg s$^{-1}$}}
\def\tbb{$kT_{\rm BB}$}
\def\rbb{$R_{\rm BB}$}

\begin{document}

\title{The role of \XMM\ in the investigation of persistent BeXRBs}

\author[1]{Nicola La Palombara*}

\author[1]{Lara Sidoli}

\author[1]{Sandro Mereghetti}

\author[2]{Gian Luca Israel}

\author[3]{Paolo Esposito}

\authormark{NICOLA LA PALOMBARA \textsc{et al}}

\address[1]{\orgdiv{INAF - IASF Milano}, \orgaddress{Via A. Corti 12, 20133 Milano, Italy}}

\address[2]{\orgdiv{INAF - Osservatorio Astronomico di Roma}, \orgaddress{via Frascati 33, 00078 Monteporzio Catone, Italy}}

\address[3]{\orgdiv{Scuola Universitaria Superiore IUSS Pavia}, \orgaddress{Palazzo del Broletto, piazza della Vittoria 15, 27100 Pavia, Italy}}

\corres{*Nicola La Palombara, \email{nicola.lapalombara@inaf.it}}

\abstract{The persistent BeXRBs are a class of High-Mass X-ray Binaries (HMXRBs), which are characterized by persistent low X-ray luminosities ($L_{\rm X} \sim 10^{34}$ erg s$^{-1}$) and wide ($P_{\rm orb} >$ 30 d), almost circular orbits. In these sources the NS is slowly rotating (with $P_{\rm spin}$ well above 100 s) and accretes matter directly from the wind of the companion Be star, without the formation of an accretion disk.

Since the '90s, when the first four members of this class were identified, several other sources of the same type have been discovered and investigated. Thanks to follow-up \textit{XMM-Newton} observations, we have verified that most of them share common spectral and timing properties, such as a pulsed fraction that does not vary with the photon energy and a hot (kT = 1-2 keV) blackbody spectral component which contributes for 20-40 \% to the total flux and has a size consistent with the NS polar cap.

Here we provide an overview of how \textit{XMM-Newton} contributed to constrain the observational properties and the current understanding of this type of sources. We also report about the first results obtained with a very recent \textit{XMM-Newton} observation of the poorly known BeXRB 4U 0728-25.}

\keywords{accretion, accretion disks -- stars: neutron -- stars: pulsars: general -- X-rays: binaries}

%%\fundingInfo{Funding info text.}

\maketitle

\section{Introduction}\label{sec1}

According to the original classification proposed by \citet{ReigRoche99}, persistent Be X-ray binaries (BeXRBs) are High Mass X-ray binaries (HMXRBs) in which a neutron star (NS) is in a wide and almost circular orbit around a Be star, with orbital period $P_{\rm orb} \gsim$ 30 days and eccentricity $e < 0.2$. In these systems the NS is characterized by a long pulse period, with $P_{\rm spin} >$ 100 s. Since the matter accretion from the Be companion star onto the NS occurs at a reduced rate, the observed luminosity is rather low, with $L_{\rm X} \sim 10^{34-35}$ \lum \citep{Pfahl+02}. This implies that commonly these systems are observed at a rather constant flux, with a limited variability (below a factor 10), and that usually they do not display significant outbursts or transient behaviour; moreover, no or very weak Iron lines at 6.4 keV are present in their spectra.

At the end of the '90s only four sources were identified as members of this class of BeXRBs, i.e. \mbox{X Persei}, \mbox{RX J0146.9+6121}, \mbox{RX J1037.5--5647}, and \mbox{RX J0440.9+4431} \citep{ReigRoche99}. Afterwards, thanks mainly to the monitoring and/or follow-up observations performed with the most recent X-ray telescopes, several other X-ray sources were identified as additional members of the persistent BeXRBs, such as e.g. \mbox{SXP 1062} \citep{Henault-Brunet+12}, \mbox{Swift J2000.6+3210} \citep{Pradhan+13}, and \mbox{XTE J1906+090} \citep{Sguera+23}. Currently almost a dozen of X-ray sources can be considered as persistent BeXRBs, and it is very likely that in the next future the members of this class of binaries will further increase.

The numerous observations and monitoring campaigns of the X-ray sky perfomed in the latest years have led to an unexpected outcome. In some rare cases, in fact, a few BeXRBs which were previously classified as persistent sources showed large outbursts, with peak luminosities even higher than $10^{37}$ \lum, as in the case of \mbox{RX J0440.9+4431} \citep{Li+24}. Apart from these transient events, these sources maintained their persistent luminosity, both before and after the observed outbursts. Therefore, these systems can still be considered as persistent sources, while the absence of any outburst as a classification criterium is no longer true.

The advent of a new generation of X-ray telescopes allowed us not only to increase the number of identified persistent BeXRBs but also to investigate in deeper detail their timing and spectral properties. Thus it was possible to obtain a rather clear and complete view of this class of sources and to point out some common properties that characterize them. In this paper we want to provide an overview of the present-day knowledge of this type of sources and, moreover, to describe the leading role played by \XMM\ \citep{Jansen+01} in this research area.

\section{X-ray observations of persistent BeXRBs}\label{sec2}

In Table~\ref{table1} we report the timing and spectral results of 15 X-ray observations of 11 persistent BeXRBs. These observations were performed along the latest 25 years with several X-ray telescopes, among which \XMM\ played a major role.

Apart from CXOU J225355.1+624336 and XTE J1906+090, all sources have a pulse period \psp\ $>$ 100 s; in particular, six of them have \psp\ $>$ 700 s. On the other hand, the other types of HMXRBs can have much smaller spin periods, down to a few milliseconds, as in the case of A0538-66 \citep{Skinner+82} and SAX J0635.2+0533 \citep{Cusumano+00}. Therefore, a long spin period of the accreting NS is one of the main common properties of this group of binary X-ray pulsars. On the other hand, although the pulsed fraction (PF) in the energy range 2-10 keV is significant for all sources, its value is strongly different depending on the specific source, varying from $\simeq$ 20 \% up to $\simeq$ 75 \%.

From the spectral point of view, in all cases their emission spectrum is well described with an absorbed power-law (PL) model, but for eight sources also an additional black-body (BB) model is necessary in order to obtain an acceptable spectral fit. Based on this emission model, in all cases the estimated luminosity (in the energy range 2-10 keV) is between $\simeq 10^{34}$ and $\simeq 10^{36}$ \lum. In all sources the spectrum is rather hard, since the PL photon index is rather low ($\Gamma < 2$), but only in four observations this parameter is $<$ 1.

In most cases the BB component has a high temperature (\tbb\ $>$ 1 keV) and a small emission region (\rbb\ $<$ 1 km). Only for one source \rbb\ $>$ 500 m, while only for two \tbb\ $>$ 2 keV. For all sources, however, the contribution of the BB component to the total flux is between $\simeq$ 20 and $\simeq$ 45 \%.

None of the observations of these sources detected an emission feature due to a possible Fe K emission line. The estimated upper limit (UL) on the Equivalent Width (EW) of this feature is always rather low, and UL$_{\rm EW} <$ 100 eV in most cases.

In this context, we note that, within the eight sources where a "\textit{hot-BB}" component was revealed, seven were observed with \XMM. In all these cases the observations performed with \XMM, thanks to its high throughput and energy resolution, were essential to assess the presence and the properties of the \textit{hot-BB} component. First of all, for several sources the \XMM\ observations allowed us to detect and investigate the pulsed emission, for the first time, also at low energies (E $<$ 2 keV). Moreover, at odds with what happened with other telescopes, only with \XMM\ it was possible to perform a detailed phase-resolved spectral analysis and, then, to prove the spectral variablity of the source along the pulse phase. This is particularly true in the cases of Swift J045106.8-694803 and 4U 0728-25, where it was possible to demonstrate that the BB component itself is clearly variable.

\section{Hot-BB component in High Mass X-Ray Binaries}\label{sec3}

\begin{figure}[t]
	\centerline{\includegraphics[width=89mm]{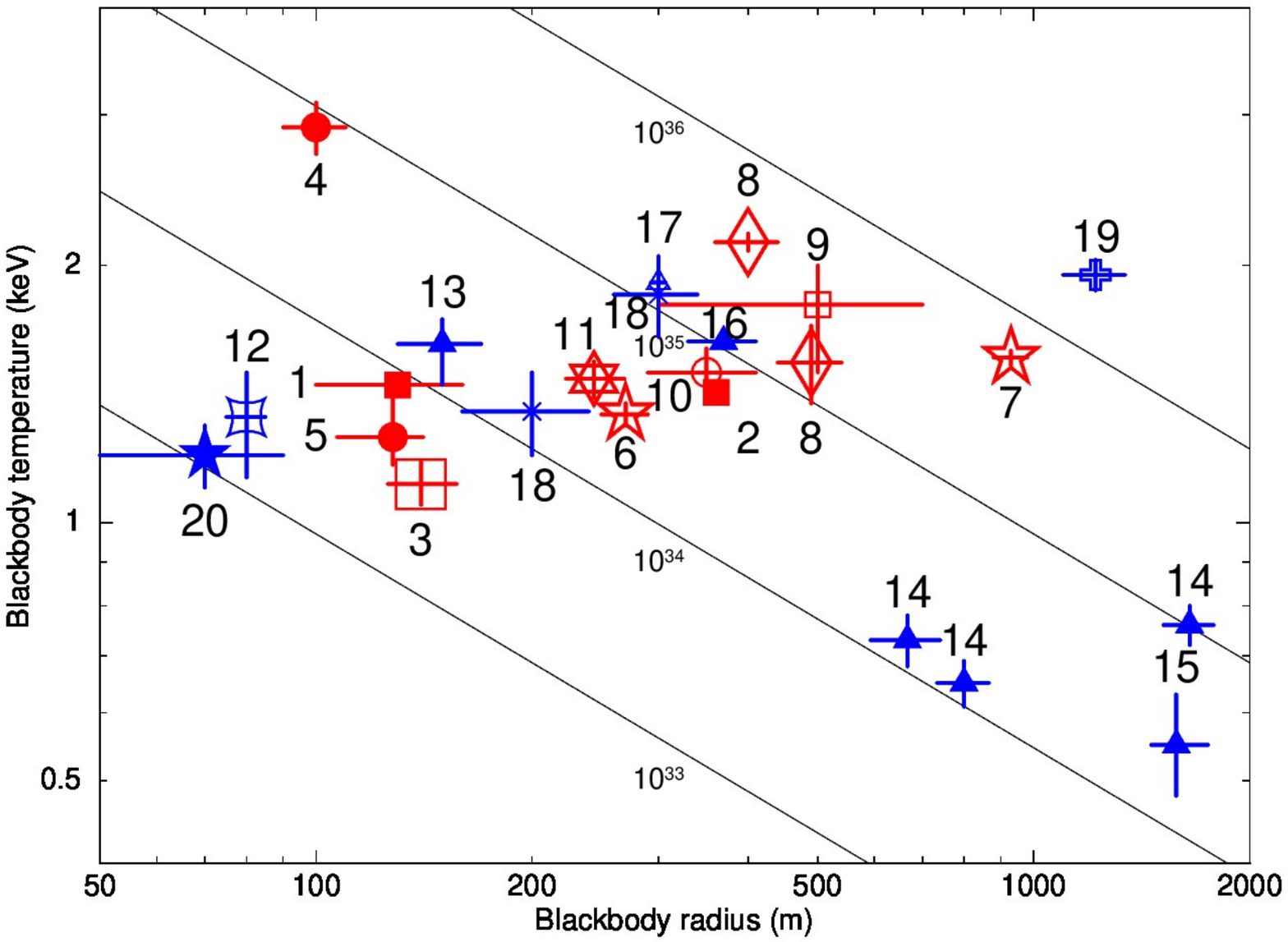}}
	\caption{Best-fit values of \rbb\ and \tbb\ of the low-luminosity and long-period pulsars with a hot BB component; red and blue symbols represent, respectively, persistent and transient accreting binary sources. The four parallel continuous lines connect the BB parameters corresponding to four different levels of luminosity (in \lum).\\
%	\begin{scriptsize}
	References:\\
	\textit{(1) X Persei \citep{Coburn+01}}\\
	\textit{(2) X Persei \citep{LaPalombara+07}}\\
	\textit{(3) RX J0146.9+6121 \citep{LaPalombara+06}}\\
	\textit{(4) RX J1037.5--5647 \citep{ReigRoche99}}\\
	\textit{(5) RX J1037.5--5647 \citep{LaPalombara+09}}\\
	\textit{(6) RX J0440.9+4431 \citep{LaPalombara+12}}\\
	\textit{(7) RX J0440.9+4431 \citep{Tsygankov+12}}\\
	\textit{(8) SXP 1062 \citep{Henault-Brunet+12}}\\
	\textit{(9) Swift J045106.8--694803 \citep{Bartlett+13}}\\
	\textit{(10) Swift J2000.6+3210 \citep{Pradhan+13}}\\
	\textit{(11) 4U 0728--25 (\textit{La Palombara et al. 2025, in preparation})}\\
	(12) 3A 0535+262 \citep{MukherjeePaul05}\\
	(13) 4U 2206+54 \citep{Masetti+04}\\
	(14) 4U 2206+54 \citep{Torrejon+04}\\
	(15) 4U 2206+54 \citep{Reig+09}\\
	(16) 4U 2206+54 \citep{Reig+12}\\
	(17) SAX J2103.5+4545 \citep{Inam+04}\\
	(18) IGR J11215--5292 \citep{Sidoli+07}\\
	(19) IGR J08408--4503 \citep{Sidoli+09}\\
	(20) XTE J1739--302 \citep{Bozzo+10}
%	\end{scriptsize}
}\label{fig1}
\end{figure}

Regarding the additional \textit{hot-BB} spectral component observed in several persistent BeXRBs, it is interesting to note that a similar component has been detected also in other types of HMXRBs. More precisely, it was observed during the low-luminosity state of a few transient HMXRBs with long pulse period. In Fig.~\ref{fig1} we show the best-fit values for the temperature and size of the BB component detected in these sources. For some sources more than one observation was performed and the estimated value of the BB temperature and/or radius can be very different in the various observations. In most cases this variation is mainly due to the different overall source luminosity in the different observations: in fact, in spite of these differences, the relative contribution of the BB component to the total source flux is almost constant (Table~\ref{table1}). Moreover, we have to take into account that observations performed with different instruments can also provide different results. Sources with reference between 1 and 11 (in \textit{italic}) are persistent BeXRBs (red symbols), while the remaining sources are transient HMXRBs (blue symbols). In the same figure we plot also parallel lines representing different levels of the total luminosity of the BB component.

We note that the transient sources reported in Fig.~\ref{fig1} were observed at luminosity levels similar to those of the persistent sources. Moreover, also the BB components detected in these sources are characterized by temperatures and radii with values fully comparable to those of the persistent sources. Finally, also in these sources the contribution of the BB component to the total source flux is in the range 20-40 \%. The only evident exception is the variable source 4U 2206+54, which in some observations showed a significantly larger and colder BB component. This can be due to the peculiar nature of this source, which very likely hosts an accreting magnetar \citep{Torrejon+18}.

These results show that not only a \textit{hot-BB} spectral component is a rather common feature of the low-luminosity and long-period accreting pulsars, but also that its parameters (temperature and size) tend to cluster in a rather narrow range of values. Therefore, it is very likely that for all these HMXRBs the origin of this component is due to the same type of physical process.

\section{Thermal excess in X-ray Binary Pulsars}\label{sec4}

The \textit{hot-BB} component observed in several low-luminosity and long-period accreting pulsars is not the only type of spectral excess (over the main continuum component) revealed in binary pulsars. In fact, a significant excess has been revealed also in several other, more luminous pulsars \citep{Woo+95,Woo+96,Yokogawa+00,LaBarbera+01,Paul+02,NaikPaul04}. Also in these cases the observed excess can be described with a thermal component, but this component is characterized by low temperature (kT $<$ 0.5 keV) and large emitting size (R $>$ 100 km), thus much larger than the NS; in most cases, this component provides a small contribution (about 5 \%) to the total source luminosity. Therefore, the properties of this component are completely different from those of the \textit{hot-BB} observed in the low-luminosity pulsars. For this reason, this type of spectral feature is called "\textit{soft excess}".

\begin{figure}[t]
	\centerline{\includegraphics[width=91mm]{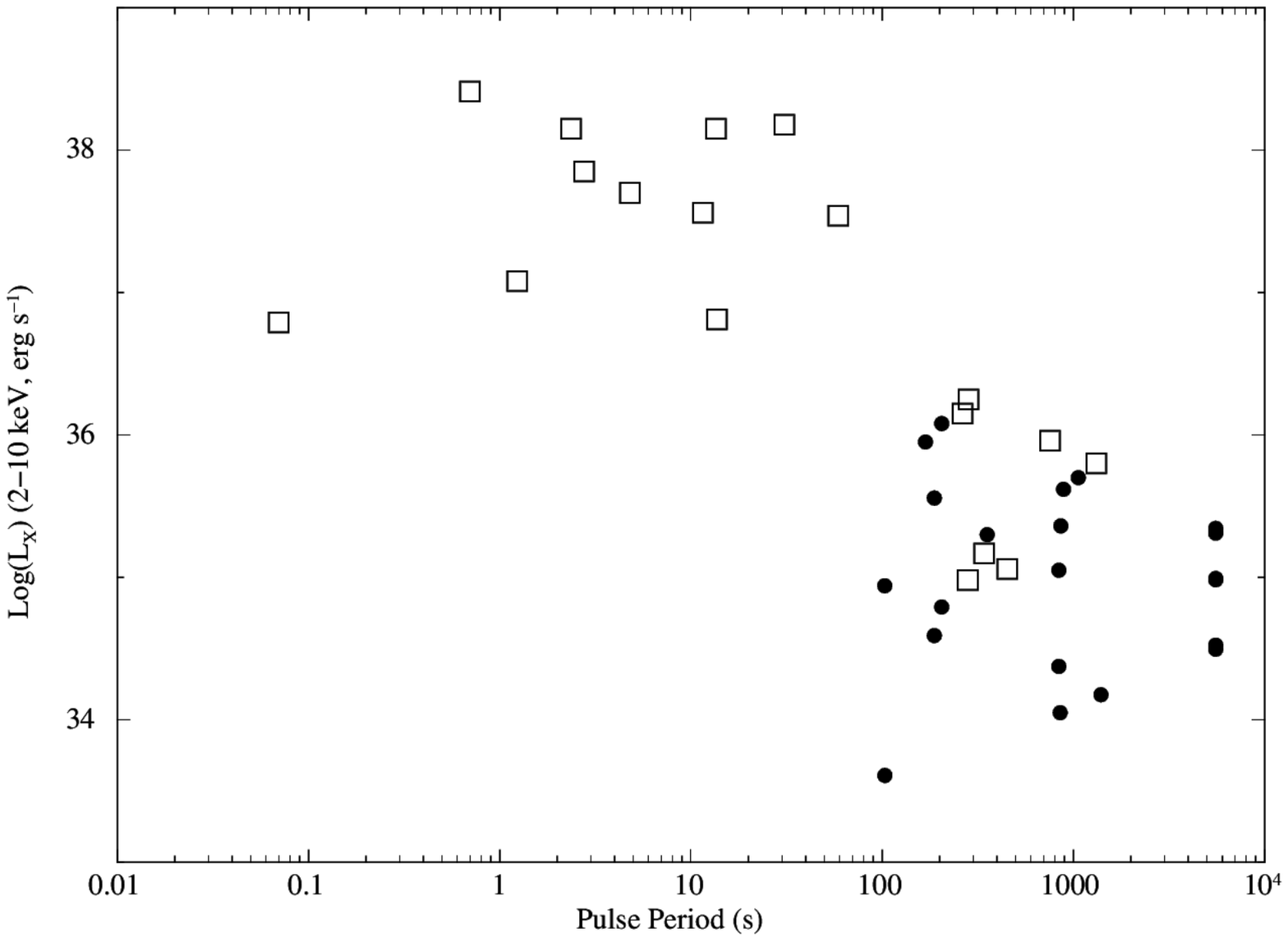}}
	\caption{X-ray luminosity and pulse period of binary pulsars showing either a \textit{soft excess} (open squares) or a \textit{hot-BB} component (filled circles).}\label{fig2}
\end{figure}

In Fig.~\ref{fig2} we report the luminosity and pulse period of all binary pulsars showing either a \textit{soft excess} (open squares) or a \textit{hot-BB} component (filled circles). It is clearly evident that they occupy two separate regions of the diagram.

\begin{itemize}
\item On one hand, there is a group of sources characterized by high luminosity (\lx\ $\gsim 10^{37}$ \lum) and short pulse period (\psp\ $<$ 100 s). Typically they are close binary systems with short orbital periods, where the matter accretion onto the NS is steady and occurs through an accretion disk, or transient pulsars observed during their outburst phase (see, e.g. \citet{LaPalombara+18}). In any case, all of them are \textit{soft excess} sources.
\item On the other hand, the remaining pulsars have lower luminosities (\lx\ $< 10^{37}$ \lum) and longer pulse periods (\psp $>$ 100 s). Most of them are in a wide orbit around the companion star and are wind-fed systems. Contrary to the previous one, in this group there are both \textit{soft excess} and \textit{hot-BB} sources.
\end{itemize}

We note that, although in the second group of pulsars there is not a clear separation of the two types of sources, on average the hot-BB component is present in the pulsars with both the lowest luminosities and the longest pulse periods. Furthermore, we emphasize that a soft excess is present only in some sources with \lx\ $\gsim 10^{35}$ \lum, while at lower luminosities only a hot BB component can be observed. Therefore, depending on the pulsar luminosity, we can identify three different groups of sources:

\begin{itemize}
\item \emph{High-luminosity} pulsars (with \lx\ $\gsim 10^{37}$ \lum), where only a soft excess has been observed
\item \emph{Intermediate-luminosity} pulsars (with $10^{35} \lsim$ \lx\ $< 10^{37}$ \lum), where both a soft excess and a hot BB can be observed
\item \emph{Low-luminosity} pulsars (with \lx\ $< 10^{35}$ \lum), where only a hot-BB component has been observed
\end{itemize}

However, we cannot exclude that, at least in some of the low-luminosity pulsars, a SE is also present. In fact, in the brighter sources where this component has been detected, it typically contributes only for a few percent to the total luminosity. Compared with the 20-40 \% contribution of the hot BB component, this means that a possible SE component could be hidden below the hot-BB one. Therefore, we caution that also in the group of low-luminosity pulsars this spectral feature could be present.

\section{Polar-cap origin of the BB component?}\label{sec5}

For all the sources with a hot BB component reported in Table~\ref{table1} the radius of the emitting region \rbb\ is much smaller than the NS size. Therefore, it is very likely that the BB emission originates from the NS surface, in agreement with the model proposed by \citet{Hickox+04}, which attributes the source of the thermal emission detected in low-luminosity accreting pulsars to the NS polar caps.

In order to verify this scenario, it is necessary to estimate the expected radius $R_{\rm col}$ of the accretion column, which is a realistic approximation of the polar cap radius, and to compare it with \rbb. To this aim, we assume $M_{\rm NS} = 1.4 M_{\odot}$, $R_{\rm NS} = 10^6$ cm, and $B_{\rm NS} = 10^{12}$ G for, respectively, the NS mass, radius, and magnetic field. From the estimated luminosity \lx\ of each source, we use the relation $\dot M$ = \lx$R_{\rm NS}$/($GM_{\rm NS}$) to derive the corresponding accretion rate. Then we estimate the magnetic dipole momentum $\mu$ = $B_{\rm NS} R_{\rm NS}^3$/2 and the magnetospheric radius $R_{\rm m}$ = \{$\mu^4$/($2GM_{\rm NS}\dot M^2$)\}$^{1/7}$. Finally, we derive the radius of the accretion column from the relation $R_{\rm col} \simeq R_{\rm NS}$($R_{\rm NS}$/$R_{\rm m}$)$^{1/2}$.

The values of $R_{\rm col}$ obtained in this way are reported in the last column of Table~\ref{table1}, where for each source also the estimated value of \rbb\ is shown. For most of the sources, the two values are consistent within a factor $\simeq$ 2; only in the case of the RXTE observation of RX J1037.5-5647 there is a significantly higher discrepancy. Taking into account the uncertainties on both the NS parameters and the source distances (within 30 \%), which affect both the estimated luminosities and BB radius, this result supports the hypothesis that, in this class of sources, the thermal excess originates at the NS polar caps, in the thermal mound at the base of the accretion column. This scenario is further strengthened by the variability of the thermal component observed in Swift J045106.8--694803 and 4U 0728--25. For these two sources, in fact, the phased-resolved spectral analysis shows that the observed spectral variability along the pulse phase cannot be attributed to a variation of only the PL component, since the simultaneous fit of the phased-resolved spectra with a constant BB component is rejected by the data. This finding can be explained by the movement of the hot spot emission from the polar cap, with respect to the observer, along the pulse phase. In the other cases there is no evidence of this type, but only of a spectral variability along the pulse phase.

\section{Conclusions}\label{sec6}
In this paper we have shown how in the latest years the class of the persistent BeXRBs has increased by including several new sources. We have also proved that \XMM\ played a leading role in this research area. Thanks to its sensitivity and spectral resolution, the \XMM\ observations of these sources enabled the assessment of their timing and spectral properties. In this way it was possible to detect a hot blackbody spectral component in most of the sources, so that now this component can be considered as an almost ubiquitous feature of the persistent BeXRBs. For most of these sources the size of the BB component is comparable with that of the NS polar cap. Therefore, it is very likely that the polar caps are the source of the observed hot BB component. This hypothesis is supported by the spectral variability observed with \XMM\ in most of the sources and, in particular, by the variability of the BB component clearly detected in two of them.

%\backmatter

\section*{Acknowledgments}

We acknowledge funding from INAF through the grant "Bando Ricerca Fondamentale INAF 2023".

%\nocite{*}% Show all bib entries - both cited and uncited; comment this line to view only cited bib entries;
\bibliography{XMM-WS2024_La-Palombara_publisher}

\clearpage

\begin{sidewaystable}[h]
\vspace{-6cm}
\caption{X-ray observations of persistent Be X-ray binaries\label{table1}}
\begin{tabular*}{\textheight}{ccccccccccc}
\toprule
\textbf{Source} & \textbf{Telescope}  & \textbf{P$_{\rm spin}$}  & \textbf{L$_{\rm X}$ (2-10 keV)} & \textbf{Photon} & \textbf{kT$_{\rm BB}$} & \textbf{R$_{\rm BB}$} & \textbf{f$_{\rm BB}$/f$_{\rm TOT}$} & \textbf{EW Fe} & \textbf{PF} & \textbf{R$_{\rm col}$} \\
\textbf{(Reference)}	&	& \textbf{(s)}  & \textbf{(erg s$^{-1}$)}  & \textbf{Index $\Gamma$}	& \textbf{(keV)}  & \textbf{(m)}  & \textbf{(\%)}  & \textbf{(eV)}  & \textbf{(\%)}  & \textbf{(m)} \\
\midrule
X Persei (1)		& RXTE	& 837  						& 2.4$\times 10^{34}$	& 1.83$\pm$0.03 			& 1.45$\pm$0.02 			& 130$\pm$30			& 24 	& $<$ 13	& 44	& 270 \\
\textbf{X Persei} (2) & \textbf{XMM-Newton}	& 839.3$\pm$0.3		& 1.1$\times 10^{35}$	& 1.48$\pm$0.02				& 1.42$^{+0.04}_{-0.02}$	& 361$\pm$3			& 39	& $<$ 100	& 41	& 330	\\
\textbf{RX J0146.9+6121} (3) & \textbf{XMM-Newton}	& 1396.1$\pm$0.3	& 1.5$\times 10^{34}$	& 1.34$^{+0.05}_{-0.06}$	& 1.11$^{+0.07}_{-0.06}$	& 140$^{+20}_{-10}$	& 24	& $<$ 150	& 69	& 250	\\
RX J1037.5-5647 (4)	& RXTE			& 860$\pm$2			& 2.3$\times 10^{35}$	& 1.84$\pm$0.06				& 2.9$\pm$0.02				& 100$\pm$10			& 45	& $<$ 65	& 53	& 370	\\
\textbf{RX J1037.5-5647} (5) & \textbf{XMM-Newton}	& 853.4$\pm$0.2		& 1.1$\times 10^{34}$	& 0.51$^{+0.17}_{-0.29}$	& 1.26$^{+0.16}_{-0.09}$	& 130$^{+10}_{-20}$	& 42	& $<$ 200	& 75	& 240	\\
\textbf{RX J0440.9+4431} (6) & \textbf{XMM-Netwon}	& 204.96$\pm$0.02	& 6.2$\times 10^{34}$	& 0.85$\pm$0.07				& 1.34$\pm$0.04				& 270$\pm$20			& 35	& $<$ 70	& 55	& 310	\\
RX J0440.9+4431 (7)	& INTEGRAL,		& 205.0$\pm$0.1		& 1.2$\times 10^{36}$	& 0.75$\pm$0.05				& 1.56$\pm$0.03				& 930$\pm$50			& 40	& $<$ 50	& 41	& 470	\\
	& Swift, RXTE	&	&	&	&	&	&	&	&	&	\\
\textbf{SXP 1062} (8)	& Chandra		& 1062				& 5.0$\times 10^{35}$	& 1.18$\pm$0.02				& 2.13$\pm$0.05				& 400$\pm$40			& 46	& -			& 20	& 410	\\
	& \textbf{XMM-Newton}	& 1062	& 5.3$\times 10^{35}$	& 0.77$\pm$0.03	& 1.54$\pm$0.16	& 490$\pm$50	& 21	& -	& 20	& 420	\\
\textbf{Swift J045106.8-694803} (9)	& \textbf{XMM-Newton}	& 168.5$\pm$0.2	& 8.9$\times 10^{35}$	& 1.4$^{+0.5}_{-0.3}$	& 1.8$^{+0.2}_{-0.3}$		& 500$\pm$200			& 41	& $<$ 300	& 41	& 450	\\
Swift J2000.6+3210 (10)	& Suzaku	& 890				& 4.2$\times 10^{35}$	& 1.04$^{+0.13}_{-0.22}$	& 1.5$\pm$0.1				& 350$\pm$60			& 42	& $<$ 15	& 41	& 400	\\
\textit{\textbf{4U 0728-25$^{a}$} (11)}	& \textit{\textbf{XMM-Newton}}		& \textit{103.301$\pm$0.005}	& \textit{8.8$\times 10^{34}$}	& \textit{1.52$\pm$0.06}		& \textit{1.48$^{+0.07}_{-0.08}$}	&	\textit{244$^{+25}_{-21}$}	& \textit{24}	& \textit{$<$ 100}	& \textit{15-23}	& \textit{320}	\\
\textbf{CXOU J225355.1+624336} (12)	& \textbf{XMM-Newton}	& 46.753$\pm$0.003	& 1.2$\times 10^{34}$	& 1.66$^{+0.14}_{-0.13}$	& -	& -	& -	& -	& 40-45	& -	\\
XTE J1906+090 (13, 14)	& Chandra, Swift,	& 89.66$\pm$0.03	& 3.1$\times 10^{34}$	& 1.3$\pm$0.3		& -	& -	& -	& -	& 21	& -	\\
	& INTEGRAL	& & & & & & & & & \\
\textbf{CXOU J182531.4-144036} (15)	& \textbf{XMM-Newton}	& 781$\pm$2	& 1.3$\times 10^{34}$	& 2.2$\pm$0.2				& -	& -	& -	& -	& 29	& -	\\
\bottomrule
\end{tabular*}
\begin{tablenotes}%%[\textheight]
\item[$^{a}$] Preliminary results
\item[References: ] (1) \citet{Coburn+01}; (2) \citet{LaPalombara+07}; (3) \citet{LaPalombara+06}; (4) \citet{ReigRoche99}; (5) \citet{LaPalombara+09}; (6) \citet{LaPalombara+12}; (7) \citet{Tsygankov+12}; (8) \citet{Henault-Brunet+12}; (9) \citet{Bartlett+13}; (10) \citet{Pradhan+13}; (11) \textit{La Palombara et al. 2025, in preparation}; (12) \citet{LaPalombara+21}; (13) \citet{Gogus+05}; (14) \citet{Sguera+23}; (15) \citet{Mason+24}
\end{tablenotes}
\end{sidewaystable}

\end{document}